\documentclass[english,reprint,a4paper,aps,pra]{revtex4-1}
\usepackage[T1]{fontenc}
\usepackage[latin9]{inputenc}
\usepackage{amsbsy}
\usepackage{graphicx}
\usepackage{esint}

\makeatletter

\providecommand{\tabularnewline}{\\}

\@ifundefined{textcolor}{}
{%
 \definecolor{BLACK}{gray}{0}
 \definecolor{WHITE}{gray}{1}
 \definecolor{RED}{rgb}{1,0,0}
 \definecolor{GREEN}{rgb}{0,1,0}
 \definecolor{BLUE}{rgb}{0,0,1}
 \definecolor{CYAN}{cmyk}{1,0,0,0}
 \definecolor{MAGENTA}{cmyk}{0,1,0,0}
 \definecolor{YELLOW}{cmyk}{0,0,1,0}
}

\providecommand \href@noop [0]{\@secondoftwo}%

\makeatother

\usepackage{babel}
\begin{document}

\title{Work relations for a system governed by Tsallis statistics }

\author{Ian J. Ford and Robert W. Eyre}

\affiliation{Department of Physics and Astronomy, University College London, Gower
Street, London WC1E 6BT, U.K.}
\begin{abstract}
We derive analogues of the Jarzynski equality and Crooks relation
to characterise the nonequilibrium work associated with changes in
the spring constant of an overdamped oscillator in a quadratically
varying spatial temperature profile. The stationary state of such
an oscillator is described by Tsallis statistics, and the work relations
for certain processes may be expressed in terms of $q$-exponentials.
We suggest that these identities might be a feature of nonequilibrium
processes in circumstances where Tsallis distributions are found.
\end{abstract}
\maketitle

\section{Introduction}

Work relations are remarkable identities associated with the behaviour
of systems undergoing nonequilibrium thermodynamic processes. As the
name suggests, they refer to the performance of mechanical work on
a system brought about by a protocol of change in external force fields
while the system exchanges heat with a coarsely specified environment.
The key point is that they hold when the thermomechanical processing
of a system takes place at an arbitrary rate, and not just in the
quasistatic limit associated with equilibrium thermodynamics. They
are therefore statements about real nonequilibrium dissipative behaviour
and have attracted attention since they provide a perspective on the
nature of entropy production \cite{Gallavotti95,Evans02,Imparato05,Crooks07,Jarworkrelations0,seifertprinciples,EvansMorriss08,Deffner08,Esposito09,Seifert12,Ford-book2013,SpinneyFordChap13}.

The identification of principles underlying the transfer of energy
in the form of heat and work motivated the initial development of
thermodynamics in the 19th century, but the construction of a framework
to justify such principles from a microscopic perspective is still
ongoing. Work relations, and the underlying fluctuation relations
or theorems, rely on rather few assumptions about the dynamics of
system components and have proved to be extremely valuable in this
regard. One of the first results in this category was developed by
Bochkov and Kuzovlev \cite{Bochkov81,Bochkov2} to describe the statistics
of nonconservative work performed on a system in an isothermal heat
bath. The Evans-Searles Fluctuation Theorem \cite{Evans93,Evans94}
provided insight into the behaviour of deterministic complex systems
driven by external mechanical and thermal interactions. A wider appreciation
of fluctuation theorems, and of the significant development that was
then underway in nonequilibrium statistical physics, followed from
the derivation of the Jarzynski equality \cite{Jarzynski97,jaroriginal1},
an elegant work relation that can be established from a number of
perspectives. Further interest in the field was generated by the Crooks
relation \cite{Crooks99}, a connection between the statistics of
work performed in two isothermal processes that are driven in a mutually
time-inverted fashion. Both relations require the system to start
out in thermal equilibrium. Experimental relevance of these results
has been demonstrated \cite{Reid04}.

We wish to explore work relations that hold for a system exposed to
an environment with a temperature gradient. The control, characterisation
and exploitation of nonisothermal conditions at the nanoscale has
received considerable attention in recent years \cite{Menges12,Millen14,Mecklenburg15}.
Thermodynamic concepts such as system temperature can remain valid
at these scales \cite{Reguera05}, though it is clear that temporal
and spatial fluctuations will be important, aspects that can be accommodated
through using stochastic thermodynamics as a theoretical description
\cite{Seifert12}. The basis of this approach is a framework of stochastic
system dynamics in which the concept of a temperature gradient enters
through a specification of the statistical properties of the environmental
noise.

We have in mind applications to experimental systems such as tweezers
\cite{Neukirch14} and nanomechanical devices \cite{Santamaria-Holek13}
that can be used to manipulate small objects such as colloidal particles.
There are challenges in establishing and controlling a temperature
profile at sub-micron spatial scales, but progress has been made in
this direction \cite{Mao05,Kumari12,Braun13}. Models of heat transport
can be used to describe the elevation of local temperature brought
about by laser absorption in a liquid medium \cite{Peterman03} and
by Joule heating in conducting nanostructures \cite{Menges12}, and
it seems that a range of thermal profiles can be established using
suitable arrangements of heat sources and sinks, together with choice
of media \cite{Braun13,Braun14}.

In such a nonisothermal environment, a system cannot achieve thermal
equilibrium, and our aim is to understand how the Jarzynski and Crooks
results need to be adapted. Work relations have been considered for
other systems that are prevented from reaching thermal equilibrium,
such as glasses \cite{Williams08b}. If the dynamics prevent or delay
evolution towards the true equilibrium, a free energy of the final
state might emerge as a valid concept, and so too might work relations.

We focus our attention on the specific case of a harmonically bound
particle in a quadratically varying thermal profile. We employ overdamped
stochastic dynamics to model the behaviour of such a particle showing
that is governed by Tsallis rather than Boltzmann statistics in the
stationary state \cite{Tsallis-book2009}. We find that by using a
suitable notation, the two work relations can be written in a very
intuitive form for certain classes of protocols. A similar generalisation
of the work relations to Tsallis statistics was proposed by Ponmurugan
\cite{Ponmurugan12} in the context of a basis in Tsallis entropy.
In contrast, our results emerge from a more standard thermodynamic
framework where Gibbs-Shannon entropy represents the uncertainty of
the state of a stochastic system. We review the nonisothermal stochastic
dynamics of systems in Section \ref{sec:Dynamical-framework}, describe
the Tsallis oscillator in Section \ref{sec:Tsallis-oscillator} and
then present analogues of the Jarzynski and Crooks results in Sections
\ref{sec:Jarzynski-equality-for} and \ref{sec:Crooks-relation-for}.
We consider whether existing experimental techniques could be used
to establish nonisothermal conditions to a degree that suits our purpose
in Section \ref{sec:Practical-considerations} before giving our conclusions
in Section \ref{sec:Conclusions}.

\section{Dynamical framework\label{sec:Dynamical-framework}}

Dynamical approaches that do not introduce noise are available \cite{Gallavotti95,Evans02},
but a stochastic framework \cite{kurchan,sekimoto1,sekimoto2} has
a number of intuitive and mathematical advantages \cite{SpinneyFordChap13}.
We consider a system that is weakly coupled to a complex environment
where the uncertainty in the interactions that take place between
them as time passes is represented by a stochastic force in the system
dynamics.

We focus attention on a particle that moves in one spatial dimension
under the influence of a confining potential, a dissipative force
proportional to the particle velocity and a stochastic force with
a strength related to the local temperature of the environment but
lacking autocorrelation. The dynamics in the overdamped limit are
specified by
\begin{equation}
\dot{x}=-\frac{1}{m\gamma}\frac{\partial\phi(x,t)}{\partial x}+\left(\frac{2k_{B}T_{r}(x)}{m\gamma}\right)^{\frac{1}{2}}\xi(t),\label{eq:100}
\end{equation}
where $x$ is the particle position, $m$ is its mass, $\phi$ is
the potential, $\gamma$ is the friction coefficient, $k_{B}$ is
the Boltzmann constant and $T_{r}$ is the temperature of the environment.
The noise $\xi$ has statistical properties $\langle\xi(t)\rangle=0$
and $\langle\xi(t)\xi(t^{\prime})\rangle=\delta(t-t^{\prime})$ where
the brackets denote an average over realisations. Equivalently, we
can express the dynamics using the following It$\bar{{\rm o}}$-rules
stochastic differential equation
\begin{equation}
dx=-\frac{\phi^{\prime}}{m\gamma}dt+\left(\frac{2k_{B}T_{r}(x)}{m\gamma}\right)^{\frac{1}{2}}dW,\label{eq:101}
\end{equation}
where $\phi^{\prime}=\partial\phi/\partial x$ and $dW$ is an increment
in a Wiener process with $\langle dW\rangle=0$ and $\langle(dW)^{2}\rangle=dt$.

The work performed on the system as it follows a trajectory (denoted
$\vec{\boldsymbol{x}}$) in an interval $0\le t\le\tau$ is
\begin{equation}
W[\vec{\boldsymbol{x}}]=\int_{0}^{\tau}\frac{\partial\phi(x(t),t))}{\partial t}dt=\int_{0}^{\tau}\dot{W}(x(t),t)dt,\label{eq:102}
\end{equation}
and for a constant environmental temperature $T_{r}=T_{0}$ it can
readily be demonstrated \cite{Jarzynski97,SpinneyFordChap13} that
\begin{equation}
\left\langle \exp\left(-W/k_{B}T_{0}\right)\right\rangle =\exp\left(-\Delta F/k_{B}T_{0}\right),\label{eq:103}
\end{equation}
where $\Delta F=F(\phi^{\tau})-F(\phi^{0})$ is the difference in
Helmholtz free energy, at temperature $T_{0}$, brought about by a
change in the potential from $\phi^{0}=\phi(x,0)$ to $\phi^{\tau}=\phi(x,\tau)$.
The brackets denote an expectation over all trajectories possible
under the dynamics, and represent either a path integration $\langle\cdot\rangle=\int d\vec{\boldsymbol{x}}{\cal P}[\vec{\boldsymbol{x}}](\cdot)$
where ${\cal P}$ is a trajectory probability density, or the average
$\langle\cdot\rangle=\int dWp(W)(\cdot)$ where $p$ is the probability
density for the work done. Eq. (\ref{eq:103}) is the celebrated Jarzynski
equality, and it holds for an arbitrary work protocol as long as the
system is initially in thermal equilibrium.

Under similar isothermal conditions and initial equilibrium, the Crooks
relation
\begin{equation}
p_{F}(W)=p_{R}(-W)\exp[(W-\Delta F)/k_{B}T_{0}],\label{eq:104}
\end{equation}
can be derived, where the suffices $F$ and $R$ refer to processes
that are time-inverted partners, in that the potential evolves over
time in process $R$ according to the reverse of the evolution in
process $F$. The free energy difference in Eq. (\ref{eq:104}) refers
to the outcome of the $F$ process.

Our purpose is to derive analogues of Eqs. (\ref{eq:103}) and (\ref{eq:104})
for work processes that take place in a nonisothermal environment
in the limit of overdamped dynamics. The appropriate form of stochastic
dynamics in these circumstances has recently received some attention
\cite{Celani12,Ge14}, particularly with regard to thermodynamic differences
arising from the neglect of housekeeping entropy production in an
overdamped treatment \cite{Ford15b}. The key point we exploit is
that with overdamped dynamics, the entropy production in the stationary
nonisothermal state vanishes, such that it takes on something of the
character of an equilibrium state, and this analogy allows us to construct
an effective free energy change associated with a process and hence
to derive work relations.

The approach we take is to recast the dynamics of Eq. (\ref{eq:100})
in an equivalent isothermal form
\begin{equation}
\dot{y}=-\frac{1}{m\gamma}\frac{\partial\Phi(y,t)}{\partial y}+\left(\frac{2k_{B}T_{0}}{m\gamma}\right)^{\frac{1}{2}}\xi(t),\label{eq:105}
\end{equation}
in terms of a stochastic variable $y(x)$ and an effective potential
$\Phi(y,t)$, and to reinterpret the work relations that emerge in
this representation. The mathematical reformulation is straightforward.
Using It$\bar{{\rm o}}$ calculus we write
\begin{eqnarray}
\!\!\!\!\!\!\! dy & = & \!\left(\!-\frac{\phi^{\prime}}{m\gamma}\frac{dy}{dx}+\frac{k_{B}T_{r}}{m\gamma}\frac{d^{2}y}{dx^{2}}\right)dt+\frac{dy}{dx}\left(\frac{2k_{B}T_{r}}{m\gamma}\right)^{\frac{1}{2}}\!\! dW,\label{eq:106}
\end{eqnarray}
and we choose $y(x)$ such that $dy/dx=(T_{0}/T_{r})^{1/2}$ and
\begin{equation}
\frac{d^{2}y}{dx^{2}}=-\frac{T_{0}^{1/2}T_{r}^{\prime}}{2T_{r}^{3/2}}\label{eq:108}
\end{equation}
where $T_{r}^{\prime}=\partial T_{r}/\partial x$, giving
\begin{equation}
dy=\left(-\frac{\phi^{\prime}}{m\gamma}-\frac{k_{B}T_{r}^{\prime}}{2m\gamma}\right)\left(\frac{T_{0}}{T_{r}}\right)^{\frac{1}{2}}dt+\left(\frac{2k_{B}T_{0}}{m\gamma}\right)^{\frac{1}{2}}dW,\label{eq:109}
\end{equation}
so that the effective potential is
\begin{eqnarray}
\Phi(y,t) & = & \int_{0}^{y}dy\left(\phi^{\prime}+\frac{1}{2}k_{B}T_{r}^{\prime}\right)\left(\frac{T_{0}}{T_{r}}\right)^{\frac{1}{2}}\nonumber \\
 & = & \int_{0}^{x(y)}dx\left(\phi^{\prime}+\frac{1}{2}k_{B}T_{r}^{\prime}\right)\left(\frac{T_{0}}{T_{r}}\right).\label{eq:110}
\end{eqnarray}
The effective isothermal work performed in this reformulation is
\begin{equation}
W_{{\rm eff}}[\vec{\boldsymbol{y}}]=\int_{0}^{\tau}dt\frac{\partial\Phi(y(t),t)}{\partial t},\label{eq:111}
\end{equation}
while the effective free energy change is $\Delta F_{{\rm eff}}=F(\Phi^{\tau})-F(\Phi^{0})$,
with $\Phi^{\tau}=\Phi(y,\tau)$ and $\Phi^{0}=\Phi(y,0)$. The Jarzynski
equality is
\begin{equation}
\int d\vec{\boldsymbol{y}}{\cal P}[\vec{\boldsymbol{y}}]\exp\left(-W_{{\rm eff}}[\vec{\boldsymbol{y}}]/k_{B}T_{0}\right)=\exp\left(-\Delta F_{{\rm eff}}/k_{B}T_{0}\right),\label{eq:112}
\end{equation}
and all that remains is to cast the left hand side as an expectation
$\int d\vec{\boldsymbol{x}}{\cal P}[\vec{\boldsymbol{x}}]\exp(-{\cal F}[\dot{W}])$
in terms of a functional ${\cal F}$ of the rate of performance of
work $\dot{W}(x,t)$.

\section{Tsallis oscillator\label{sec:Tsallis-oscillator}}

We illustrate how Eq. (\ref{eq:112}) can be interpreted for a model
of 1-d particle motion in a harmonic confining potential and a temperature
profile that varies quadratically with position measured from the
centre of the potential. We consider $\phi(x,t)=\phi_{\kappa}=\kappa(t)x^{2}/2$
and
\begin{equation}
T_{r}(x)=T_{0}\left(1+\frac{\kappa_{T}x^{2}}{2k_{B}T_{0}}\right),\label{eq:113}
\end{equation}
where $\kappa_{T}$ is a parameter with the same dimensions as the
spring constant $\kappa$. This system is of interest since the Fokker-Planck
equation for the pdf $P(x,t)$ corresponding to the dynamics of Eq.
(\ref{eq:100}) is
\begin{equation}
\frac{\partial P}{\partial t}=\frac{\partial}{\partial x}\left(\frac{\kappa x}{m\gamma}P+\frac{k_{B}}{m\gamma}\frac{\partial(T_{r}P)}{\partial x}\right),\label{eq:114}
\end{equation}
for which the stationary solution is
\begin{equation}
P_{{\rm st}}^{\kappa}(x)=\left[\frac{\kappa_{T}}{2\pi k_{B}T_{0}}\right]^{\frac{1}{2}}\frac{\Gamma\left(1+\kappa/\kappa_{T}\right)}{\Gamma\left(\frac{1}{2}+\kappa/\kappa_{T}\right)}\left(1+\frac{\kappa_{T}x^{2}}{2k_{B}T_{0}}\right)^{-\frac{\kappa}{\kappa_{T}}-1},\label{eq:115}
\end{equation}
namely a Tsallis distribution, proportional to a $q$-exponential
function defined by
\begin{equation}
{\rm e}_{q}(z)=\left(1+(1-q)z\right)^{1/(1-q)},\label{eq:116}
\end{equation}
here with $q=(\kappa+2\kappa_{T})/(\kappa+\kappa_{T})$ and $z=-(\kappa+\kappa_{T})x^{2}/2k_{B}T_{0}$.
Such distributions have received attention because of the ubiquity
of experimental data that take this form \cite{Tsallis-book2009}.
They contrast with the more usual gaussians characteristic of equilibrium
situations, which correspond to the limit $q\to1$, or $\kappa_{T}\to0$
in this case. The work relations that we derive will refer to processes
initiated when the system is in such a stationary state.

The effective potential $\Phi(y,t)=\Phi_{\kappa}(y)$ can be found
in closed form for this model. We have
\begin{eqnarray}
\Phi_{\kappa}(y) & = & \int_{0}^{x(y)}dx\left(\kappa x+\frac{1}{2}\kappa_{T}x\right)\left(1+\frac{\kappa_{T}x^{2}}{2k_{B}T_{0}}\right)^{-1}\nonumber \\
 & = & \left(\kappa+\frac{1}{2}\kappa_{T}\right)\frac{k_{B}T_{0}}{\kappa_{T}}\ln\left(1+\frac{\kappa_{T}x^{2}}{2k_{B}T_{0}}\right).\label{eq:117}
\end{eqnarray}
From this we can write
\begin{equation}
W_{{\rm eff}}=\int_{0}^{\tau}dt\dot{\kappa}\frac{k_{B}T_{0}}{\kappa_{T}}\ln\left(1+\frac{\kappa_{T}x^{2}}{2k_{B}T_{0}}\right),\label{eq:118}
\end{equation}
so that the functional that appears in the Jarzynski equality is
\begin{equation}
{\cal F}[\dot{W}]=\int_{0}^{\tau}dt\dot{W}\frac{2}{\kappa_{T}x^{2}}\ln\left(1+\frac{\kappa_{T}x^{2}}{2k_{B}T_{0}}\right),\label{eq:119}
\end{equation}
using $\dot{W}=\dot{\kappa}x^{2}/2$. This becomes clearer to interpret
when written in the form
\begin{equation}
{\cal F}[\dot{W}]=\int_{0}^{\tau}dt\frac{\dot{W}}{k_{B}T_{0}}\left[\frac{T_{0}}{(T_{r}-T_{0})}\ln\left(\frac{T_{r}}{T_{0}}\right)\right].\label{eq:120}
\end{equation}
The factor in square brackets is unity for an isothermal bath, which
can be demonstrated by writing $T_{r}=T_{0}+\epsilon(x)$, such that
for small $\epsilon$ we have $[T_{0}/(T_{r}-T_{0})]\ln(T_{r}/T_{0})\approx(T_{0}/\epsilon)[\epsilon/T_{0}+O(\epsilon^{2})]=1+O(\epsilon)$
after which we take $\epsilon\to0$, and the factor is less than or
equal to unity for nonisothermal conditions. For an isothermal bath
we would write ${\cal F}=W/k_{B}T_{0}$.

For a given $x(t)$, the functional ${\cal F}$ of the associated
work rate can be determined and $\langle\exp(-{\cal F})\rangle$ set
equal to $\exp(-[F(\Phi_{\kappa(\tau)})-F(\Phi_{\kappa(0)})]/k_{B}T_{0})$
with
\begin{eqnarray}
\exp[-F(\Phi_{\kappa})/k_{B}T_{0}]\propto\int dy\exp(-\Phi_{\kappa}(y)/k_{B}T_{0})\nonumber \\
=\int dy\left[\cosh[(\kappa_{T}/2k_{B}T_{0})^{1/2}y]\right]^{-2\kappa/\kappa_{T}-1}\nonumber \\
=\pi^{\frac{1}{2}}\frac{\Gamma\left(\frac{1}{2}+\kappa/\kappa_{T}\right)}{\Gamma\left(1+\kappa/\kappa_{T}\right)},\qquad\qquad\label{eq:121}
\end{eqnarray}
noting that $x=(2k_{B}T_{0}/\kappa_{T})^{1/2}\sinh[(\kappa_{T}/2k_{B}T_{0})^{1/2}y]$
so that $\Phi_{\kappa}(y)/k_{B}T_{0}=\left[2\kappa/\kappa_{T}+1\right]\ln\cosh[(\kappa_{T}/2k_{B}T_{0})^{1/2}y]$.
In short, the Jarzynski equality for this system is
\begin{eqnarray}
\left\langle \exp-\int_{0}^{\tau}dt\frac{\dot{W}}{k_{B}T_{0}}\left[\frac{T_{0}}{(T_{r}-T_{0})}\ln\left(\frac{T_{r}}{T_{0}}\right)\right]\right\rangle \nonumber \\
=\frac{\Gamma\left(\frac{1}{2}+\kappa(\tau)/\kappa_{T}\right)}{\Gamma\left(1+\kappa(\tau)/\kappa_{T}\right)}\frac{\Gamma\left(1+\kappa(0)/\kappa_{T}\right)}{\Gamma\left(\frac{1}{2}+\kappa(0)/\kappa_{T}\right)}.\label{eq:122}
\end{eqnarray}

\section{Jarzynski equality for multistep processes\label{sec:Jarzynski-equality-for}}

\begin{figure}
\begin{centering}
\includegraphics[width=1\columnwidth]{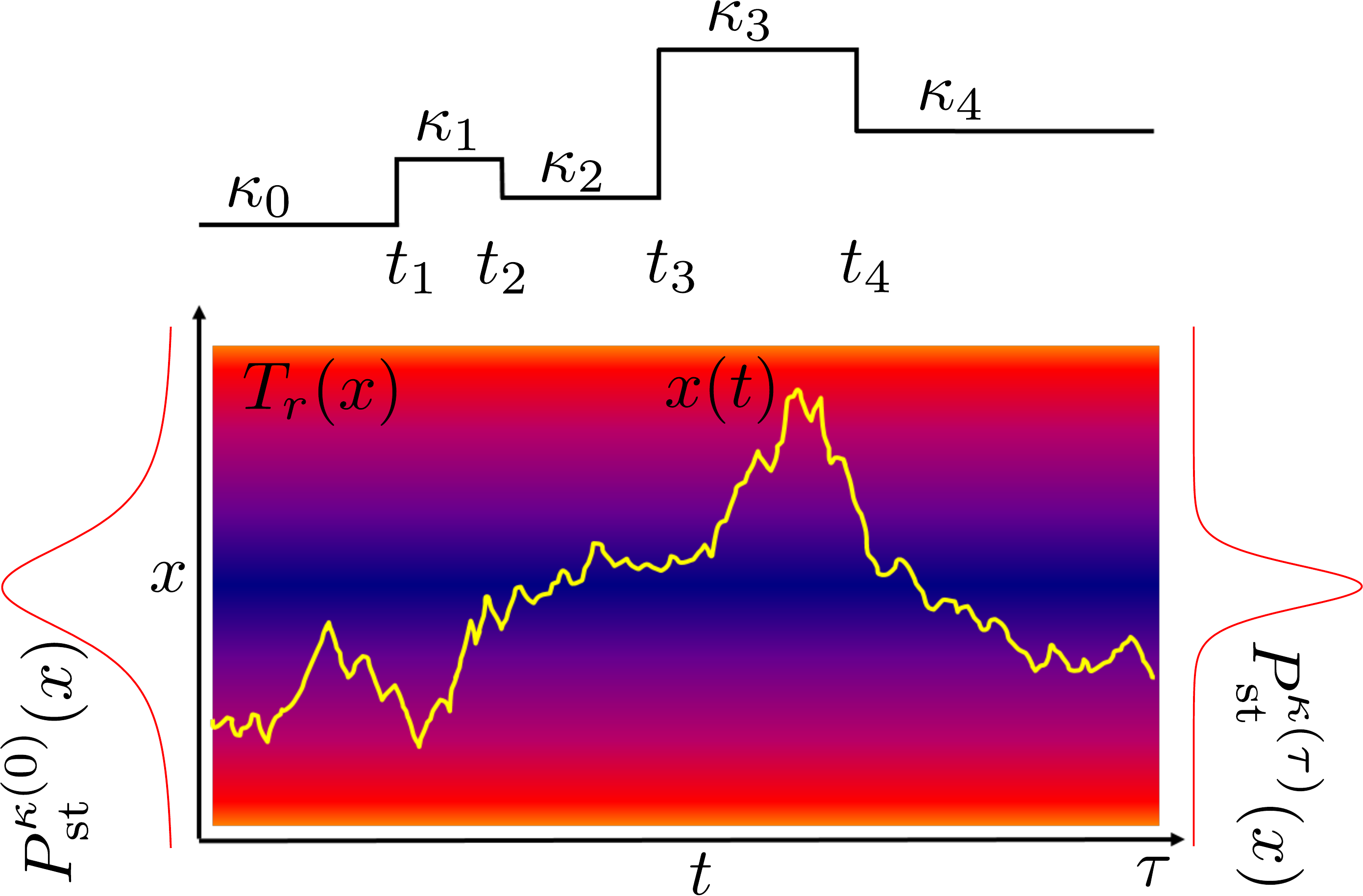}
\par\end{centering}

\protect\caption{Illustration of a particle trajectory $x(t)$ in a nonisothermal environment
(indicated by background colour) as the spring constant changes in
a sequence of steps. The initial distribution of positions takes the
stationary Tsallis form $P_{{\rm st}}^{\kappa(0)}$. At time $\tau$
the distribution need not take the stationary form $P_{{\rm st}}^{\kappa(\tau)}$,
shown for comparison. \label{fig:Trajectory.}}
\end{figure}

We focus our attention on work processes consisting of a set of $N$
abrupt changes in the spring constant, i.e. shifts $\kappa_{i-1}\to\kappa_{i}$
at times $t=t_{i}$ for $i=1$ to $N$, with $\kappa_{0}=\kappa(0)$
and $\kappa_{N}=\kappa(\tau)$. The protocol of spring constant shifts
is illustrated in Figure \ref{fig:Trajectory.} together with a representation
of a stochastic trajectory $x(t)$ in the presence of a background
temperature profile $T_{r}(x)$. Tsallis distributions $P_{{\rm st}}^{\kappa}$
associated with stationary states at the beginning and end of the
period are also shown: the system begins in a stationary state, but
it need not end the process in such a condition. Note that a form
of ergodic consistency holds here in the sense that the probability
density is nowhere zero during the evolution: the stochastic dynamics
can take the system from any initial position to any final position.

For such a multistep process the functional ${\cal F}$ becomes
\begin{equation}
{\cal F}(\{\kappa_{i}\},\{x_{i}\})=\sum_{i=1}^{N}\frac{(\kappa_{i}-\kappa_{i-1})}{\kappa_{T}}\ln\left(1+\frac{\kappa_{T}x_{i}^{2}}{2k_{B}T_{0}}\right),\label{eq:123}
\end{equation}
where $x_{i}=x(t_{i})$. Hence
\begin{eqnarray}
\exp(-{\cal F}) & = & \prod_{i=1}^{N}\left(1+\frac{\kappa_{T}x_{i}^{2}}{2k_{B}T_{0}}\right)^{-(\kappa_{i}-\kappa_{i-1})/\kappa_{T}}\nonumber \\
 & = & \prod_{i=1}^{N}{\rm e}_{q_{i}}\left(-\frac{(\kappa_{i}-\kappa_{i-1})x_{i}^{2}}{2k_{B}T_{0}}\right),\label{eq:124}
\end{eqnarray}
where $q_{i}=1+\kappa_{T}/(\kappa_{i}-\kappa_{i-1})$. This leads
to the result
\begin{equation}
\left\langle \prod_{i=1}^{N}{\rm e}_{q_{i}}\left[-\frac{\Delta W_{i}}{k_{B}T_{0}}\right]\right\rangle =\frac{\Gamma\left(\frac{1}{2}+\kappa_{N}/\kappa_{T}\right)}{\Gamma\left(1+\kappa_{N}/\kappa_{T}\right)}\frac{\Gamma\left(1+\kappa_{0}/\kappa_{T}\right)}{\Gamma\left(\frac{1}{2}+\kappa_{0}/\kappa_{T}\right)},\label{eq:125}
\end{equation}
where $\Delta W_{i}=(\kappa_{i}-\kappa_{i-1})x_{i}^{2}/2$ is the
work associated with the $i$th step change in spring constant, such
that the total work done is $W=\sum_{1}^{N}\Delta W_{i}$. The notation
now makes very apparent the resemblance to the isothermal Jarzynski
equality in Eq. (\ref{eq:103}). The brackets here refer to averaging
according to $\langle\cdot\rangle=\int\prod_{0}^{N}(\cdot)dx_{i}p_{{\rm T}}(x_{0},x_{1},\cdots,x_{N})$
where $x_{0}=x(0)$ and $p_{{\rm T}}$ is the probability that the
dynamics generate a discrete trajectory $\{x_{i}\}$. If $\kappa_{T}\to0$
the $q$-exponentials tend towards ordinary exponentials and the $\Gamma$
functions can be represented using Stirling's approximation such that
the right hand side of Eq. (\ref{eq:125}) reduces to $(\kappa_{0}/\kappa_{N})^{1/2}$,
as expected.

\begin{figure}
\begin{centering}
\includegraphics[width=1\columnwidth]{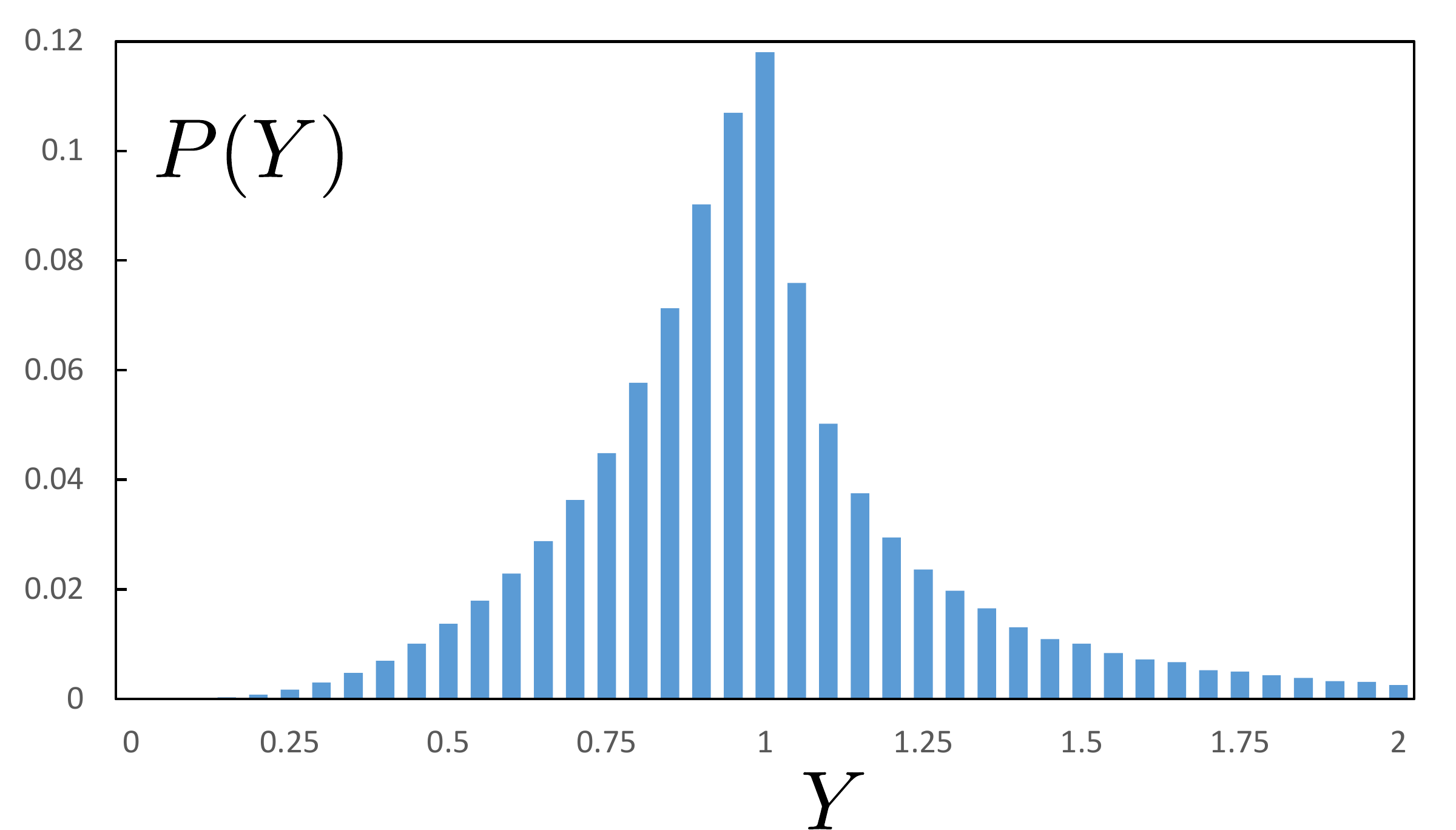}
\par\end{centering}

\protect\caption{Distribution of values of $Y=\prod_{i=1}^{4}{\rm e}_{q_{i}}\left(-\Delta W_{i}/k_{B}T_{0}\right)$
for a cyclic process with $\kappa_{0}=1$, $\kappa_{1}=1.5$, $\kappa_{2}=2$,
$\kappa_{3}=0.5$, and $\kappa_{4}=1$, with $t_{2}-t_{1}=t_{3}-t_{2}=t_{4}-t_{3}=0.1$,
and $m=\gamma=k_{B}=T_{0}=\kappa_{T}=1$, obtained from $10^{5}$
realisations with a timestep of $10^{-4}$. The mean of the distribution
is 0.999 with statistical uncertainty of 0.001, verifying the analogue
Jarzynski equality (\ref{eq:125}) for this process. \label{fig:Jarzynski}}
\end{figure}

It is straightforward to verify the identity (\ref{eq:125}) in simple
cases such as a process consisting of a step up from $\kappa_{0}$
to $\kappa_{1}$ at time $t_{1}=0$ followed by a step back down to
$\kappa_{2}=\kappa_{0}$ at $t_{2}=\tau$. We select initial positions
according to the stationary distribution $P_{{\rm st}}^{\kappa_{0}}(x_{0})$
given in Eq. (\ref{eq:115}) with a given $\kappa_{T}$, and solve
the stochastic dynamics in Eq. (\ref{eq:101}) numerically in order
to evaluate $x_{1}$ and $x_{2}$ and hence $\Delta W_{1}$ and $\Delta W_{2}$.
The expectation of ${\rm e}_{q_{1}}\left(-\Delta W_{1}/k_{B}T_{0}\right){\rm e}_{q_{2}}\left(-\Delta W_{2}/k_{B}T_{0}\right)$
over a number of realisations can be shown to be consistent with unity,
within statistical uncertainty, as indicated in Table \ref{tab:Demonstration-of-the}.
An illustration of the distribution of $\prod_{i=1}^{N}{\rm e}_{q_{i}}\left(-\Delta W_{i}/k_{B}T_{0}\right)$
for a more complex cyclic process is shown in Figure \ref{fig:Jarzynski}.

\begin{table}
\begin{centering}
\begin{tabular}{|c|c|c|c|}
\hline
$\kappa_{1}/\kappa_{0}$ & $\kappa_{T}/\kappa_{0}$ & $q$ & LHS of Eq. (\ref{eq:125})\tabularnewline
\hline
\hline
1.01 & 1.005 & 1.5 & $1.0000\pm0.0001$\tabularnewline
\hline
1.05 & 1.025 & 1.49 & $1.000\pm0.0011$\tabularnewline
\hline
1.10 & 1.05 & 1.49 & $1.000\pm0.0017$\tabularnewline
\hline
1.20 & 1.10 & 1.48 & $1.000\pm0.0021$\tabularnewline
\hline
1.40 & 1.20 & 1.46 & $1.001\pm0.0072$\tabularnewline
\hline
1.75 & 1.375 & 1.44 & $1.00\pm0.012$\tabularnewline
\hline
2.0 & 1.0 & 1.33 & $1.00\pm0.025$\tabularnewline
\hline
4.0 & 2.5 & 1.38 & $1.00\pm0.085$\tabularnewline
\hline
6.0 & 3.5 & 1.37 & $1.0\pm0.12$\tabularnewline
\hline
\end{tabular}
\par\end{centering}

\protect\caption{Demonstration of the Jarzynski equality for the Tsallis oscillator
for a range of work processes consisting of a step up and down in
spring constant (from $\kappa_{0}$ to $\kappa_{1}$ and back again)
separated by a period $\tau=4$, with a variety of spatial temperature
profiles specified by $\kappa_{T}$, in each case starting from a
Tsallis distribution of position characterised by $q$, with $m=\gamma=k_{B}=T_{0}=\kappa_{0}=1$.
Results are based on 1000 numerical realisations for each set of parameters,
using a timestep of $10^{-5}$. Errors in the mean are obtained assuming
each realisation is an independent sample. The left hand side (LHS)
of Eq. (\ref{eq:125}) is expected to equal unity. \label{tab:Demonstration-of-the}}
\end{table}

\section{Crooks relation for a step process\label{sec:Crooks-relation-for}}

\begin{figure}
\begin{centering}
\includegraphics[width=1\columnwidth]{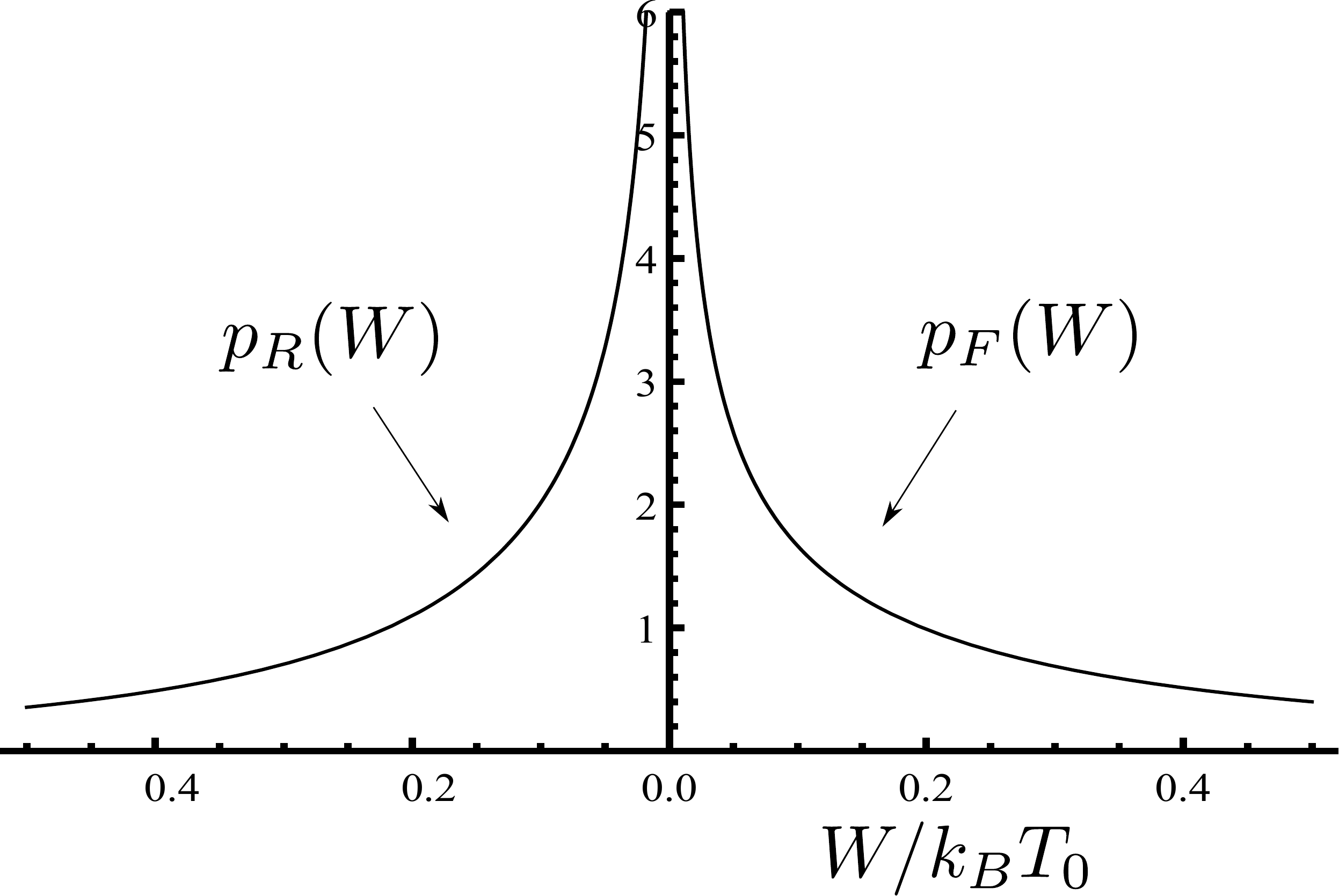}
\par\end{centering}

\protect\caption{Probability distributions of work for process $F$ consisting of a
step up from $\kappa_{0}=\kappa_{T}$ to $\kappa_{1}=2\kappa_{T}$
(positive values of $W$), and process $R$ specified by a step down
from $\kappa_{1}$ to $\kappa_{0}$ (negative values of $W$), both
starting from a stationary state. The mean work for the $F$ process
is $k_{B}T_{0}$ and for the $R$ process it is $-k_{B}T_{0}/3$.
\label{fig:stuff}}
\end{figure}

Our next task is to determine the analogue of the Crooks relation
for a Tsallis oscillator. The simplest demonstration involves an $F$
process consisting of a step up in spring constant from $\kappa_{0}$
to $\kappa_{1}\ge\kappa_{0}$ starting from a stationary state, such
that the corresponding $R$ process is the step down from $\kappa_{1}$
to $\kappa_{0}$. It is straightforward to evaluate the statistics
of work performed from the statistics of position prior to the step
change. We write
\begin{equation}
p_{F}(W)=P_{{\rm st}}^{\kappa_{0}}(x_{0})\frac{dx_{0}}{dW},\label{eq:126}
\end{equation}
with $W=(\kappa_{1}-\kappa_{0})x_{0}^{2}/2\ge0$ such that $dx_{0}/dW=[2(\kappa_{1}-\kappa_{0})W]^{-1/2}$
and so with appropriate normalisation we have
\begin{eqnarray}
p_{F}(W) & = & \left[\frac{\kappa_{T}}{\pi k_{B}T_{0}}\frac{1}{(\kappa_{1}-\kappa_{0})W}\right]^{\frac{1}{2}}\frac{\Gamma\left(1+\kappa_{0}/\kappa_{T}\right)}{\Gamma\left(\frac{1}{2}+\kappa_{0}/\kappa_{T}\right)}\nonumber \\
 & \times & {\rm e}_{q_{0F}}\left(-\frac{(\kappa_{0}+\kappa_{T})W}{(\kappa_{1}-\kappa_{0})k_{B}T_{0}}\right)\Theta(W),\label{eq:127}
\end{eqnarray}
with $q_{0F}=(\kappa_{0}+2\kappa_{T})/(\kappa_{0}+\kappa_{T})$ and
where $\Theta$ is the Heaviside function. Similarly we can determine
the distribution of work $W=(\kappa_{0}-\kappa_{1})x_{0}^{2}/2\le0$
for the reverse process starting from the stationary state with $\kappa=\kappa_{1}$:
\begin{eqnarray}
p_{R}(W) & = & \left[\frac{\kappa_{T}}{\pi k_{B}T_{0}}\frac{1}{(\kappa_{0}-\kappa_{1})W}\right]^{\frac{1}{2}}\frac{\Gamma\left(1+\kappa_{1}/\kappa_{T}\right)}{\Gamma\left(\frac{1}{2}+\kappa_{1}/\kappa_{T}\right)}\nonumber \\
 & \times & {\rm e}_{q_{0R}}\left(-\frac{(\kappa_{1}+\kappa_{T})W}{(\kappa_{0}-\kappa_{1})k_{B}T_{0}}\right)\Theta(-W),\label{eq:128}
\end{eqnarray}
with $q_{0R}=(\kappa_{1}+2\kappa_{T})/(\kappa_{1}+\kappa_{T})$. These
two distributions are illustrated for the case of $\kappa_{0}/\kappa_{T}=1$
and $\kappa_{1}/\kappa_{T}=2$ in Figure \ref{fig:stuff}. The analogue
Crooks relation is then
\begin{eqnarray}
\frac{p_{F}(W)}{p_{R}(-W)} & = & \frac{\Gamma\left(\frac{1}{2}+\kappa_{1}/\kappa_{T}\right)}{\Gamma\left(1+\kappa_{1}/\kappa_{T}\right)}\frac{\Gamma\left(1+\kappa_{0}/\kappa_{T}\right)}{\Gamma\left(\frac{1}{2}+\kappa_{0}/\kappa_{T}\right)}\nonumber \\
 &  & \times\left(1+\frac{\kappa_{T}W}{(\kappa_{1}-\kappa_{0})k_{B}T_{0}}\right)^{-\frac{\kappa_{0}}{\kappa_{T}}-1}\nonumber \\
 &  & \times\left(1-\frac{\kappa_{T}W}{(\kappa_{0}-\kappa_{1})k_{B}T_{0}}\right)^{\frac{\kappa_{1}}{\kappa_{T}}+1},\label{eq:129}
\end{eqnarray}
defined in this case for $W\ge0$, which may be written in the form
\begin{eqnarray}
\frac{p_{F}(W)}{p_{R}(-W)} & = & \exp(-[F(\Phi_{\kappa_{1}})-F(\Phi_{\kappa_{0}})]/k_{B}T_{0})\nonumber \\
 & \times & \left(1+\frac{\kappa_{T}W}{(\kappa_{1}-\kappa_{0})k_{B}T_{0}}\right)^{(\kappa_{1}-\kappa_{0})/\kappa_{T}},\label{eq:130}
\end{eqnarray}
and the final factor can be written as ${\rm e}_{q_{R}}(W/k_{B}T_{0})$
with $q_{R}=1-\kappa_{T}/(\kappa_{1}-\kappa_{0})$ or $\left[{\rm e}_{q_{F}}(-W/k_{B}T_{0})\right]^{-1}$
with $q_{F}=1+\kappa_{T}/(\kappa_{1}-\kappa_{0})$. Once again, the
notation makes it quite apparent that the Crooks relation for the
pair of processes under isothermal conditions, in the form of Eq.
(\ref{eq:104}), is recovered when $q_{F,R}\to1$ and $\Phi_{\kappa}\to\phi_{\kappa}$
as $\kappa_{T}\to0$.

We can further conclude from these distributions that the mean work
performed in the $F$ process $\langle W\rangle_{F}=\int Wp_{F}(W)dW$
is given by $(\kappa_{1}-\kappa_{0})k_{B}T_{0}/(2\kappa_{0}-\kappa_{T})$,
as long as $\kappa_{0}\ge\kappa_{T}/2$, and the mean work for the
$R$ process is $-(\kappa_{1}-\kappa_{0})k_{B}T_{0}/(2\kappa_{1}-\kappa_{T})$,
assuming that $\kappa_{1}\ge\kappa_{T}/2$. For a process consisting
of a step up, the establishment of a stationary state, and a step
down, the mean work would therefore be
\begin{eqnarray}
\langle W\rangle_{F}+\langle W\rangle_{R} & = & \frac{2(\kappa_{1}-\kappa_{0})^{2}k_{B}T_{0}}{(2\kappa_{0}-\kappa_{T})(2\kappa_{1}-\kappa_{T})},\label{eq:131}
\end{eqnarray}
which is never negative, and furthermore tends towards a known result
\cite{Reid04} for such a process in the isothermal limit as $\kappa_{T}\to0$.

\begin{figure}
\begin{centering}
\includegraphics[width=1\columnwidth]{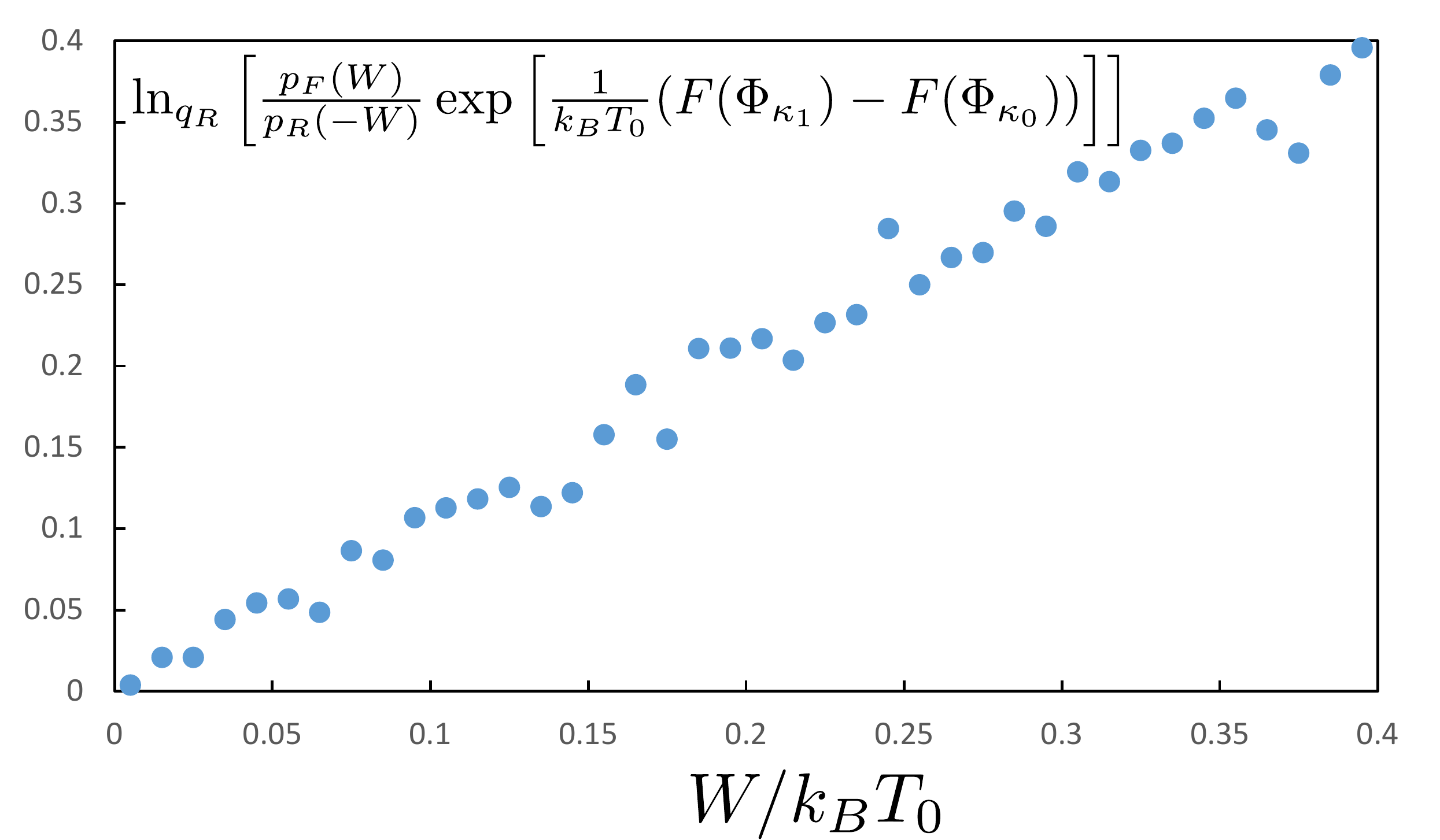}
\par\end{centering}

\protect\caption{Verification of the analogue Crooks relation Eq. (\ref{eq:130}) for
a step up from $\kappa_{0}=\kappa_{T}$ to $\kappa_{1}=2\kappa_{T}$
and a step down from $\kappa_{1}$ to $\kappa_{0}$, based on $10^{6}$
particle trajectories with a timestep of $10^{-4}$, with $m=\gamma=k_{B}=T_{0}=\kappa_{T}=1$.\label{fig:Crooks}}
\end{figure}

As an illustration of a possible experimental verification of the
analogue Crooks relation, we have generated trajectories according
to Eq. (\ref{eq:100}) for a step up from $\kappa_{0}/\kappa_{T}=1$
to $\kappa_{1}/\kappa_{T}=2$, and the corresponding step down, both
starting from the appropriate stationary state, in order to compute
the work distributions. A plot of $\ln_{q_{R}}[p_{F}(W/k_{B}T_{0})/p_{R}(-W/k_{B}T_{0})\exp([F(\Phi_{\kappa_{1}})-F(\Phi_{\kappa_{0}})]/k_{B}T_{0})]$
against $W/k_{B}T_{0}$ should be a straight line with a gradient
of unity, where $\ln_{q}(z)=(z^{1-q}-1)/(1-q)$ is the $q$-logarithm
\cite{Tsallis-book2009}. Such an outcome is apparent in Figure \ref{fig:Crooks}.

\section{Practical considerations\label{sec:Practical-considerations}}

The work relations will depart significantly from the form taken under
isothermal conditions when $\kappa_{T}/\kappa\sim1$ and we should
consider whether current methods for establishing a thermal profile
in a particle trap can provide such circumstances. Spring constants
for particle trapping with optical tweezers in a liquid medium are
typically of the order of $10^{-2}$ pN/nm \cite{Rohrbach05,Kumari12}
 and if we adopt this value for $\kappa_{T}$ in Eq. (\ref{eq:113}),
the thermal gradient $\kappa_{T}x/k_{B}$ at a typical displacement
of 10 nm from the trap centre would be about $10$ K/nm. This is not
an unattainable gradient in a solid system \cite{Menges12,Santamaria-Holek13}
but could pose experimental challenges for a fluid system.

 In contrast, an approximately quadratic temperature profile over
a spatial scale of microns with a peak thermal gradient of about 10
K/$\mu$m can be established by laser illumination of a patterned
nanostructure immersed in a fluid medium \cite{Braun13}. A particle
can be confined within such a profile by thermophoresis (thermal diffusion)
\cite{Piazza08} or potentially by optical methods as well. Thermophoretic
confinement provides much weaker trapping, with a spring constant
of order $10^{-5}$ pN/nm. The effect is equivalent to the presence
of a potential given in dimensionless form by $\phi(x)/k_{B}T=S_{T}\Delta T_{r}(x)$,
where $T$ is an effective constant ambient temperature and $S_{T}$
is the Soret coefficient that compares the strength of thermal diffusion,
in response to a temperature gradient, with that of ordinary diffusion
down a density gradient. For a quadratic temperature profile the confinement
is harmonic with an effective spring constant $\kappa=TS_{T}\kappa_{T}$
and strongly nonisothermal conditions will therefore be established
if $TS_{T}\sim1$. Such a thermophoretic trap was demonstrated in
\cite{Braun13} using 200 nm polystyrene spheres with $S_{T}\sim3$
K$^{-1}$, and hence $\kappa_{T}\sim10^{-3}\kappa$, but by trapping
smaller species the Soret coefficient could potentially be reduced
to $10^{-2}$ K$^{-1}$. There remains the challenge of ensuring that
such a trap is strong enough to retain a particle for a long enough
time to perform measurements, and compromises will need to be made
between the desired temperature profile and such requirements. Nevertheless,
it seems possible that with further technological development the
conditions of interest might be attainable.

\section{Conclusions\label{sec:Conclusions}}

Work relations exist for systems that are maintained away from thermal
equilibrium by constraints, and a number of such cases have been discussed
\cite{hatanosasa,IFThousekeeping,SpinneyFord12b}. The situation we
have explored here is of interest since the system in question, when
described using overdamped dynamics, is governed by Tsallis statistics
in its stationary state. Such statistical properties have been noted
in a variety of physical systems \cite{Tsallis-book2009}, and contrast
with the more usual gaussian statistics of thermal equilibrium. The
analogues of the Jarzynski equality and Crooks relation for such a
system, constructed for certain processes, take a particularly transparent
form when we make use of the $q$-exponential functions associated
with Tsallis statistics.

Stochastic entropy production can provide the conceptual underpinning
of work relations in nonisothermal as well as isothermal conditions,
employing the usual Gibbs-Shannon form rather than the Tsallis entropy.
We believe that experimental verification of these identities might
be possible in small systems where the appropriate nonisothermal environment
can be established and controlled.

\section*{Acknowledgements}

We thank Janet Anders for her helpful support and comments on this
work.


\begin{thebibliography}{45}%
\makeatletter
\providecommand \@ifxundefined [1]{%
 \@ifx{#1\undefined}
}%
\providecommand \@ifnum [1]{%
 \ifnum #1\expandafter \@firstoftwo
 \else \expandafter \@secondoftwo
 \fi
}%
\providecommand \@ifx [1]{%
 \ifx #1\expandafter \@firstoftwo
 \else \expandafter \@secondoftwo
 \fi
}%
\providecommand \natexlab [1]{#1}%
\providecommand \enquote  [1]{``#1''}%
\providecommand \bibnamefont  [1]{#1}%
\providecommand \bibfnamefont [1]{#1}%
\providecommand \citenamefont [1]{#1}%
\providecommand \href@noop [0]{\@secondoftwo}%
\providecommand \href [0]{\begingroup \@sanitize@url \@href}%
\providecommand \@href[1]{\@@startlink{#1}\@@href}%
\providecommand \@@href[1]{\endgroup#1\@@endlink}%
\providecommand \@sanitize@url [0]{\catcode `\\12\catcode `\$12\catcode
  `\&12\catcode `\#12\catcode `\^12\catcode `\_12\catcode `\%12\relax}%
\providecommand \@@startlink[1]{}%
\providecommand \@@endlink[0]{}%
\providecommand \url  [0]{\begingroup\@sanitize@url \@url }%
\providecommand \@url [1]{\endgroup\@href {#1}{\urlprefix }}%
\providecommand \urlprefix  [0]{URL }%
\providecommand \Eprint [0]{\href }%
\providecommand \doibase [0]{http://dx.doi.org/}%
\providecommand \selectlanguage [0]{\@gobble}%
\providecommand \bibinfo  [0]{\@secondoftwo}%
\providecommand \bibfield  [0]{\@secondoftwo}%
\providecommand \translation [1]{[#1]}%
\providecommand \BibitemOpen [0]{}%
\providecommand \bibitemStop [0]{}%
\providecommand \bibitemNoStop [0]{.\EOS\space}%
\providecommand \EOS [0]{\spacefactor3000\relax}%
\providecommand \BibitemShut  [1]{\csname bibitem#1\endcsname}%
\let\auto@bib@innerbib\@empty
\bibitem [{\citenamefont {Gallavotti}\ and\ \citenamefont
  {Cohen}(1995)}]{Gallavotti95}%
  \BibitemOpen
  \bibfield  {author} {\bibinfo {author} {\bibfnamefont {G.}~\bibnamefont
  {Gallavotti}}\ and\ \bibinfo {author} {\bibfnamefont {E.~G.~D.}\ \bibnamefont
  {Cohen}},\ }\href@noop {} {\bibfield  {journal} {\bibinfo  {journal} {Phys.
  Rev. Lett.}\ }\textbf {\bibinfo {volume} {74}},\ \bibinfo {pages} {2694}
  (\bibinfo {year} {1995})}\BibitemShut {NoStop}%
\bibitem [{\citenamefont {Evans}\ and\ \citenamefont
  {Searles}(2002)}]{Evans02}%
  \BibitemOpen
  \bibfield  {author} {\bibinfo {author} {\bibfnamefont {D.~J.}\ \bibnamefont
  {Evans}}\ and\ \bibinfo {author} {\bibfnamefont {D.~J.}\ \bibnamefont
  {Searles}},\ }\href@noop {} {\bibfield  {journal} {\bibinfo  {journal} {Adv.
  Phys.}\ }\textbf {\bibinfo {volume} {51}},\ \bibinfo {pages} {1529} (\bibinfo
  {year} {2002})}\BibitemShut {NoStop}%
\bibitem [{\citenamefont {Imparato}\ and\ \citenamefont
  {Peliti}(2005)}]{Imparato05}%
  \BibitemOpen
  \bibfield  {author} {\bibinfo {author} {\bibfnamefont {A.}~\bibnamefont
  {Imparato}}\ and\ \bibinfo {author} {\bibfnamefont {L.}~\bibnamefont
  {Peliti}},\ }\href@noop {} {\bibfield  {journal} {\bibinfo  {journal} {Phys.
  Rev. E}\ }\textbf {\bibinfo {volume} {72}},\ \bibinfo {pages} {046114}
  (\bibinfo {year} {2005})}\BibitemShut {NoStop}%
\bibitem [{\citenamefont {Crooks}\ and\ \citenamefont
  {Jarzynski}(2007)}]{Crooks07}%
  \BibitemOpen
  \bibfield  {author} {\bibinfo {author} {\bibfnamefont {G.~E.}\ \bibnamefont
  {Crooks}}\ and\ \bibinfo {author} {\bibfnamefont {C.}~\bibnamefont
  {Jarzynski}},\ }\href@noop {} {\bibfield  {journal} {\bibinfo  {journal}
  {Phys. Rev. E}\ }\textbf {\bibinfo {volume} {75}},\ \bibinfo {pages} {021116}
  (\bibinfo {year} {2007})}\BibitemShut {NoStop}%
\bibitem [{\citenamefont {Jarzynski}(2007)}]{Jarworkrelations0}%
  \BibitemOpen
  \bibfield  {author} {\bibinfo {author} {\bibfnamefont {C.}~\bibnamefont
  {Jarzynski}},\ }\href@noop {} {\bibfield  {journal} {\bibinfo  {journal} {C.
  R. Physique}\ }\textbf {\bibinfo {volume} {8}},\ \bibinfo {pages} {495 }
  (\bibinfo {year} {2007})}\BibitemShut {NoStop}%
\bibitem [{\citenamefont {Seifert}(2008)}]{seifertprinciples}%
  \BibitemOpen
  \bibfield  {author} {\bibinfo {author} {\bibfnamefont {U.}~\bibnamefont
  {Seifert}},\ }\href@noop {} {\bibfield  {journal} {\bibinfo  {journal} {Eur.
  Phys. J. B}\ }\textbf {\bibinfo {volume} {64}},\ \bibinfo {pages} {423}
  (\bibinfo {year} {2008})}\BibitemShut {NoStop}%
\bibitem [{\citenamefont {Evans}\ and\ \citenamefont
  {Morriss}(2008)}]{EvansMorriss08}%
  \BibitemOpen
  \bibfield  {author} {\bibinfo {author} {\bibfnamefont {D.~J.}\ \bibnamefont
  {Evans}}\ and\ \bibinfo {author} {\bibfnamefont {G.}~\bibnamefont
  {Morriss}},\ }\href@noop {} {\emph {\bibinfo {title} {Statistical Mechanics
  of Nonequilibrium Liquids, Second Edition}}}\ (\bibinfo  {publisher}
  {Cambridge University Press},\ \bibinfo {year} {2008})\BibitemShut {NoStop}%
\bibitem [{\citenamefont {Deffner}\ and\ \citenamefont
  {Lutz}(2008)}]{Deffner08}%
  \BibitemOpen
  \bibfield  {author} {\bibinfo {author} {\bibfnamefont {S.}~\bibnamefont
  {Deffner}}\ and\ \bibinfo {author} {\bibfnamefont {E.}~\bibnamefont {Lutz}},\
  }\href@noop {} {\bibfield  {journal} {\bibinfo  {journal} {Phys. Rev. E}\
  }\textbf {\bibinfo {volume} {77}},\ \bibinfo {pages} {021128} (\bibinfo
  {year} {2008})}\BibitemShut {NoStop}%
\bibitem [{\citenamefont {Esposito}\ \emph {et~al.}(2009)\citenamefont
  {Esposito}, \citenamefont {Harbola},\ and\ \citenamefont
  {Mukamel}}]{Esposito09}%
  \BibitemOpen
  \bibfield  {author} {\bibinfo {author} {\bibfnamefont {M.}~\bibnamefont
  {Esposito}}, \bibinfo {author} {\bibfnamefont {U.}~\bibnamefont {Harbola}}, \
  and\ \bibinfo {author} {\bibfnamefont {S.}~\bibnamefont {Mukamel}},\
  }\href@noop {} {\bibfield  {journal} {\bibinfo  {journal} {Rev. Mod. Phys.}\
  }\textbf {\bibinfo {volume} {81}},\ \bibinfo {pages} {1665} (\bibinfo {year}
  {2009})}\BibitemShut {NoStop}%
\bibitem [{\citenamefont {Seifert}(2012)}]{Seifert12}%
  \BibitemOpen
  \bibfield  {author} {\bibinfo {author} {\bibfnamefont {U.}~\bibnamefont
  {Seifert}},\ }\href@noop {} {\bibfield  {journal} {\bibinfo  {journal} {Rep.
  Prog. Phys.}\ }\textbf {\bibinfo {volume} {75}},\ \bibinfo {pages} {126001}
  (\bibinfo {year} {2012})}\BibitemShut {NoStop}%
\bibitem [{\citenamefont {{I.J. Ford}}(2013)}]{Ford-book2013}%
  \BibitemOpen
  \bibfield  {author} {\bibinfo {author} {\bibnamefont {{I.J. Ford}}},\
  }\href@noop {} {\emph {\bibinfo {title} {Statistical Physics: an entropic
  approach}}}\ (\bibinfo  {publisher} {Wiley},\ \bibinfo {year}
  {2013})\BibitemShut {NoStop}%
\bibitem [{\citenamefont {{R.E. Spinney and I.J.
  Ford}}(2013)}]{SpinneyFordChap13}%
  \BibitemOpen
  \bibfield  {author} {\bibinfo {author} {\bibnamefont {{R.E. Spinney and I.J.
  Ford}}},\ }in\ \href {http://arXiv:1201.6381v1} {\emph {\bibinfo {booktitle}
  {Nonequilibrium Statistical Physics of Small Systems: Fluctuation Relations
  and Beyond}}},\ \bibinfo {editor} {edited by\ \bibinfo {editor}
  {\bibfnamefont {R.~J.}\ \bibnamefont {Klages}}, \bibinfo {editor}
  {\bibfnamefont {W.}~\bibnamefont {Just}}, \ and\ \bibinfo {editor}
  {\bibfnamefont {C.}~\bibnamefont {Jarzynski}}}\ (\bibinfo  {publisher}
  {Wiley-VCH, Weinheim, ISBN 978-3-527-41094-1},\ \bibinfo {year}
  {2013})\BibitemShut {NoStop}%
\bibitem [{\citenamefont {Bochkov}\ and\ \citenamefont
  {Kuzovlev}(1981{\natexlab{a}})}]{Bochkov81}%
  \BibitemOpen
  \bibfield  {author} {\bibinfo {author} {\bibfnamefont {G.}~\bibnamefont
  {Bochkov}}\ and\ \bibinfo {author} {\bibfnamefont {Y.}~\bibnamefont
  {Kuzovlev}},\ }\href@noop {} {\bibfield  {journal} {\bibinfo  {journal}
  {Physica A}\ }\textbf {\bibinfo {volume} {106}},\ \bibinfo {pages} {443 }
  (\bibinfo {year} {1981}{\natexlab{a}})}\BibitemShut {NoStop}%
\bibitem [{\citenamefont {Bochkov}\ and\ \citenamefont
  {Kuzovlev}(1981{\natexlab{b}})}]{Bochkov2}%
  \BibitemOpen
  \bibfield  {author} {\bibinfo {author} {\bibfnamefont {G.}~\bibnamefont
  {Bochkov}}\ and\ \bibinfo {author} {\bibfnamefont {Y.}~\bibnamefont
  {Kuzovlev}},\ }\href@noop {} {\bibfield  {journal} {\bibinfo  {journal}
  {Physica A}\ }\textbf {\bibinfo {volume} {106}},\ \bibinfo {pages} {480 }
  (\bibinfo {year} {1981}{\natexlab{b}})}\BibitemShut {NoStop}%
\bibitem [{\citenamefont {Evans}\ \emph {et~al.}(1993)\citenamefont {Evans},
  \citenamefont {Cohen},\ and\ \citenamefont {Morriss}}]{Evans93}%
  \BibitemOpen
  \bibfield  {author} {\bibinfo {author} {\bibfnamefont {D.~J.}\ \bibnamefont
  {Evans}}, \bibinfo {author} {\bibfnamefont {E.~G.~D.}\ \bibnamefont {Cohen}},
  \ and\ \bibinfo {author} {\bibfnamefont {G.~P.}\ \bibnamefont {Morriss}},\
  }\href@noop {} {\bibfield  {journal} {\bibinfo  {journal} {Phys. Rev. Lett.}\
  }\textbf {\bibinfo {volume} {71}},\ \bibinfo {pages} {2401} (\bibinfo {year}
  {1993})}\BibitemShut {NoStop}%
\bibitem [{\citenamefont {Evans}\ and\ \citenamefont
  {Searles}(1994)}]{Evans94}%
  \BibitemOpen
  \bibfield  {author} {\bibinfo {author} {\bibfnamefont {D.~J.}\ \bibnamefont
  {Evans}}\ and\ \bibinfo {author} {\bibfnamefont {D.~J.}\ \bibnamefont
  {Searles}},\ }\href {\doibase 10.1103/PhysRevE.50.1645} {\bibfield  {journal}
  {\bibinfo  {journal} {Phys. Rev. E}\ }\textbf {\bibinfo {volume} {50}},\
  \bibinfo {pages} {1645} (\bibinfo {year} {1994})}\BibitemShut {NoStop}%
\bibitem [{\citenamefont {Jarzynski}(1997{\natexlab{a}})}]{Jarzynski97}%
  \BibitemOpen
  \bibfield  {author} {\bibinfo {author} {\bibfnamefont {C.}~\bibnamefont
  {Jarzynski}},\ }\href@noop {} {\bibfield  {journal} {\bibinfo  {journal}
  {Phys. Rev. Lett.}\ }\textbf {\bibinfo {volume} {78}},\ \bibinfo {pages}
  {2690} (\bibinfo {year} {1997}{\natexlab{a}})}\BibitemShut {NoStop}%
\bibitem [{\citenamefont {Jarzynski}(1997{\natexlab{b}})}]{jaroriginal1}%
  \BibitemOpen
  \bibfield  {author} {\bibinfo {author} {\bibfnamefont {C.}~\bibnamefont
  {Jarzynski}},\ }\href@noop {} {\bibfield  {journal} {\bibinfo  {journal}
  {Phys. Rev. E}\ }\textbf {\bibinfo {volume} {56}},\ \bibinfo {pages} {5018}
  (\bibinfo {year} {1997}{\natexlab{b}})}\BibitemShut {NoStop}%
\bibitem [{\citenamefont {Crooks}(1999)}]{Crooks99}%
  \BibitemOpen
  \bibfield  {author} {\bibinfo {author} {\bibfnamefont {G.~E.}\ \bibnamefont
  {Crooks}},\ }\href@noop {} {\bibfield  {journal} {\bibinfo  {journal} {Phys.
  Rev. E}\ }\textbf {\bibinfo {volume} {60}},\ \bibinfo {pages} {2721}
  (\bibinfo {year} {1999})}\BibitemShut {NoStop}%
\bibitem [{\citenamefont {Reid}\ \emph {et~al.}(2004)\citenamefont {Reid},
  \citenamefont {Carberry}, \citenamefont {Wang}, \citenamefont {Sevick},
  \citenamefont {Evans},\ and\ \citenamefont {Searles}}]{Reid04}%
  \BibitemOpen
  \bibfield  {author} {\bibinfo {author} {\bibfnamefont {J.~C.}\ \bibnamefont
  {Reid}}, \bibinfo {author} {\bibfnamefont {D.~M.}\ \bibnamefont {Carberry}},
  \bibinfo {author} {\bibfnamefont {G.~M.}\ \bibnamefont {Wang}}, \bibinfo
  {author} {\bibfnamefont {E.~M.}\ \bibnamefont {Sevick}}, \bibinfo {author}
  {\bibfnamefont {D.~J.}\ \bibnamefont {Evans}}, \ and\ \bibinfo {author}
  {\bibfnamefont {D.~J.}\ \bibnamefont {Searles}},\ }\href@noop {} {\bibfield
  {journal} {\bibinfo  {journal} {Phys. Rev. E}\ }\textbf {\bibinfo {volume}
  {70}},\ \bibinfo {pages} {016111} (\bibinfo {year} {2004})}\BibitemShut
  {NoStop}%
\bibitem [{\citenamefont {Menges}\ \emph {et~al.}(2012)\citenamefont {Menges},
  \citenamefont {Riel}, \citenamefont {Stemmer},\ and\ \citenamefont
  {Gotsmann}}]{Menges12}%
  \BibitemOpen
  \bibfield  {author} {\bibinfo {author} {\bibfnamefont {F.}~\bibnamefont
  {Menges}}, \bibinfo {author} {\bibfnamefont {H.}~\bibnamefont {Riel}},
  \bibinfo {author} {\bibfnamefont {A.}~\bibnamefont {Stemmer}}, \ and\
  \bibinfo {author} {\bibfnamefont {B.}~\bibnamefont {Gotsmann}},\ }\href@noop
  {} {\bibfield  {journal} {\bibinfo  {journal} {Nano. Lett.}\ }\textbf
  {\bibinfo {volume} {12}},\ \bibinfo {pages} {596} (\bibinfo {year}
  {2012})}\BibitemShut {NoStop}%
\bibitem [{\citenamefont {Millen}\ \emph {et~al.}(2014)\citenamefont {Millen},
  \citenamefont {Deesuwan}, \citenamefont {Barker},\ and\ \citenamefont
  {Anders}}]{Millen14}%
  \BibitemOpen
  \bibfield  {author} {\bibinfo {author} {\bibfnamefont {J.}~\bibnamefont
  {Millen}}, \bibinfo {author} {\bibfnamefont {T.}~\bibnamefont {Deesuwan}},
  \bibinfo {author} {\bibfnamefont {P.}~\bibnamefont {Barker}}, \ and\ \bibinfo
  {author} {\bibfnamefont {J.}~\bibnamefont {Anders}},\ }\href@noop {}
  {\bibfield  {journal} {\bibinfo  {journal} {Nature Nanotechnology}\ }\textbf
  {\bibinfo {volume} {9}},\ \bibinfo {pages} {425} (\bibinfo {year}
  {2014})}\BibitemShut {NoStop}%
\bibitem [{\citenamefont {Mecklenburg}\ \emph {et~al.}(2015)\citenamefont
  {Mecklenburg}, \citenamefont {Hubbard}, \citenamefont {White}, \citenamefont
  {Dhall}, \citenamefont {Cronin}, \citenamefont {Aloni},\ and\ \citenamefont
  {Regan}}]{Mecklenburg15}%
  \BibitemOpen
  \bibfield  {author} {\bibinfo {author} {\bibfnamefont {M.}~\bibnamefont
  {Mecklenburg}}, \bibinfo {author} {\bibfnamefont {W.~A.}\ \bibnamefont
  {Hubbard}}, \bibinfo {author} {\bibfnamefont {E.~R.}\ \bibnamefont {White}},
  \bibinfo {author} {\bibfnamefont {R.}~\bibnamefont {Dhall}}, \bibinfo
  {author} {\bibfnamefont {S.~B.}\ \bibnamefont {Cronin}}, \bibinfo {author}
  {\bibfnamefont {S.}~\bibnamefont {Aloni}}, \ and\ \bibinfo {author}
  {\bibfnamefont {B.~C.}\ \bibnamefont {Regan}},\ }\href@noop {} {\bibfield
  {journal} {\bibinfo  {journal} {Science}\ }\textbf {\bibinfo {volume}
  {347}},\ \bibinfo {pages} {629} (\bibinfo {year} {2015})}\BibitemShut
  {NoStop}%
\bibitem [{\citenamefont {Reguera}\ \emph {et~al.}(2005)\citenamefont
  {Reguera}, \citenamefont {Rubi},\ and\ \citenamefont {Vilar}}]{Reguera05}%
  \BibitemOpen
  \bibfield  {author} {\bibinfo {author} {\bibfnamefont {D.}~\bibnamefont
  {Reguera}}, \bibinfo {author} {\bibfnamefont {J.~M.}\ \bibnamefont {Rubi}}, \
  and\ \bibinfo {author} {\bibfnamefont {J.~M.~G.}\ \bibnamefont {Vilar}},\
  }\href@noop {} {\bibfield  {journal} {\bibinfo  {journal} {J. Phys. Chem. B}\
  }\textbf {\bibinfo {volume} {109}},\ \bibinfo {pages} {21502} (\bibinfo
  {year} {2005})}\BibitemShut {NoStop}%
\bibitem [{\citenamefont {Neukirch}\ and\ \citenamefont
  {Vamivakas}(2014)}]{Neukirch14}%
  \BibitemOpen
  \bibfield  {author} {\bibinfo {author} {\bibfnamefont {L.}~\bibnamefont
  {Neukirch}}\ and\ \bibinfo {author} {\bibfnamefont {A.~N.}\ \bibnamefont
  {Vamivakas}},\ }\href@noop {} {\bibfield  {journal} {\bibinfo  {journal}
  {Contemporary Physics}\ }\textbf {\bibinfo {volume} {56}},\ \bibinfo {pages}
  {48} (\bibinfo {year} {2014})}\BibitemShut {NoStop}%
\bibitem [{\citenamefont {Santamaria-Holek}\ \emph {et~al.}(2013)\citenamefont
  {Santamaria-Holek}, \citenamefont {Reguera},\ and\ \citenamefont
  {Rubi}}]{Santamaria-Holek13}%
  \BibitemOpen
  \bibfield  {author} {\bibinfo {author} {\bibfnamefont {I.}~\bibnamefont
  {Santamaria-Holek}}, \bibinfo {author} {\bibfnamefont {D.}~\bibnamefont
  {Reguera}}, \ and\ \bibinfo {author} {\bibfnamefont {J.~M.}\ \bibnamefont
  {Rubi}},\ }\href@noop {} {\bibfield  {journal} {\bibinfo  {journal} {J. Phys.
  Chem. C}\ }\textbf {\bibinfo {volume} {117}},\ \bibinfo {pages} {3109}
  (\bibinfo {year} {2013})}\BibitemShut {NoStop}%
\bibitem [{\citenamefont {Mao}\ \emph {et~al.}(2005)\citenamefont {Mao},
  \citenamefont {Arias-Gonzaliez}, \citenamefont {Smith}, \citenamefont {{I.
  Tinoco Jr.}},\ and\ \citenamefont {Bustamante}}]{Mao05}%
  \BibitemOpen
  \bibfield  {author} {\bibinfo {author} {\bibfnamefont {H.}~\bibnamefont
  {Mao}}, \bibinfo {author} {\bibfnamefont {J.~R.}\ \bibnamefont
  {Arias-Gonzaliez}}, \bibinfo {author} {\bibfnamefont {S.~B.}\ \bibnamefont
  {Smith}}, \bibinfo {author} {\bibnamefont {{I. Tinoco Jr.}}}, \ and\ \bibinfo
  {author} {\bibfnamefont {C.}~\bibnamefont {Bustamante}},\ }\href@noop {}
  {\bibfield  {journal} {\bibinfo  {journal} {Biophys. J.}\ }\textbf {\bibinfo
  {volume} {89}},\ \bibinfo {pages} {1308} (\bibinfo {year}
  {2005})}\BibitemShut {NoStop}%
\bibitem [{\citenamefont {Kumari}\ \emph {et~al.}(2012)\citenamefont {Kumari},
  \citenamefont {Dharmadhikari}, \citenamefont {Dharmadhikari}, \citenamefont
  {Basu}, \citenamefont {Sharma},\ and\ \citenamefont {Mathur}}]{Kumari12}%
  \BibitemOpen
  \bibfield  {author} {\bibinfo {author} {\bibfnamefont {P.}~\bibnamefont
  {Kumari}}, \bibinfo {author} {\bibfnamefont {J.~A.}\ \bibnamefont
  {Dharmadhikari}}, \bibinfo {author} {\bibfnamefont {A.~K.}\ \bibnamefont
  {Dharmadhikari}}, \bibinfo {author} {\bibfnamefont {H.}~\bibnamefont {Basu}},
  \bibinfo {author} {\bibfnamefont {S.}~\bibnamefont {Sharma}}, \ and\ \bibinfo
  {author} {\bibfnamefont {D.}~\bibnamefont {Mathur}},\ }\href@noop {}
  {\bibfield  {journal} {\bibinfo  {journal} {Optics Express}\ }\textbf
  {\bibinfo {volume} {20}},\ \bibinfo {pages} {4645} (\bibinfo {year}
  {2012})}\BibitemShut {NoStop}%
\bibitem [{\citenamefont {Braun}\ and\ \citenamefont {Cichos}(2013)}]{Braun13}%
  \BibitemOpen
  \bibfield  {author} {\bibinfo {author} {\bibfnamefont {M.}~\bibnamefont
  {Braun}}\ and\ \bibinfo {author} {\bibfnamefont {F.}~\bibnamefont {Cichos}},\
  }\href@noop {} {\bibfield  {journal} {\bibinfo  {journal} {ACS Nano}\
  }\textbf {\bibinfo {volume} {7}},\ \bibinfo {pages} {11200} (\bibinfo {year}
  {2013})}\BibitemShut {NoStop}%
\bibitem [{\citenamefont {Peterman}\ \emph {et~al.}(2003)\citenamefont
  {Peterman}, \citenamefont {Gittes},\ and\ \citenamefont
  {Schmidt}}]{Peterman03}%
  \BibitemOpen
  \bibfield  {author} {\bibinfo {author} {\bibfnamefont {E.~J.}\ \bibnamefont
  {Peterman}}, \bibinfo {author} {\bibfnamefont {F.}~\bibnamefont {Gittes}}, \
  and\ \bibinfo {author} {\bibfnamefont {C.~F.}\ \bibnamefont {Schmidt}},\
  }\href@noop {} {\bibfield  {journal} {\bibinfo  {journal} {Biophys. J.}\
  }\textbf {\bibinfo {volume} {84}},\ \bibinfo {pages} {1308} (\bibinfo {year}
  {2003})}\BibitemShut {NoStop}%
\bibitem [{\citenamefont {Braun}\ \emph {et~al.}(2014)\citenamefont {Braun},
  \citenamefont {W{\"u}rger},\ and\ \citenamefont {Cichos}}]{Braun14}%
  \BibitemOpen
  \bibfield  {author} {\bibinfo {author} {\bibfnamefont {M.}~\bibnamefont
  {Braun}}, \bibinfo {author} {\bibfnamefont {A.}~\bibnamefont {W{\"u}rger}}, \
  and\ \bibinfo {author} {\bibfnamefont {F.}~\bibnamefont {Cichos}},\
  }\href@noop {} {\bibfield  {journal} {\bibinfo  {journal} {Phys. Chem. Chem.
  Phys.}\ }\textbf {\bibinfo {volume} {16}},\ \bibinfo {pages} {15207}
  (\bibinfo {year} {2014})}\BibitemShut {NoStop}%
\bibitem [{\citenamefont {Williams}\ \emph {et~al.}(2008)\citenamefont
  {Williams}, \citenamefont {Searles},\ and\ \citenamefont
  {Evans}}]{Williams08b}%
  \BibitemOpen
  \bibfield  {author} {\bibinfo {author} {\bibfnamefont {S.~R.}\ \bibnamefont
  {Williams}}, \bibinfo {author} {\bibfnamefont {D.~J.}\ \bibnamefont
  {Searles}}, \ and\ \bibinfo {author} {\bibfnamefont {D.~J.}\ \bibnamefont
  {Evans}},\ }\href@noop {} {\bibfield  {journal} {\bibinfo  {journal} {J.
  Chem. Phys.}\ }\textbf {\bibinfo {volume} {129}},\ \bibinfo {pages} {134504}
  (\bibinfo {year} {2008})}\BibitemShut {NoStop}%
\bibitem [{\citenamefont {Tsallis}(2009)}]{Tsallis-book2009}%
  \BibitemOpen
  \bibfield  {author} {\bibinfo {author} {\bibfnamefont {C.}~\bibnamefont
  {Tsallis}},\ }\href@noop {} {\emph {\bibinfo {title} {Introduction to
  Nonextensive Statistical Mechanics}}}\ (\bibinfo  {publisher} {Springer},\
  \bibinfo {year} {2009})\BibitemShut {NoStop}%
\bibitem [{\citenamefont {Ponmurugan}()}]{Ponmurugan12}%
  \BibitemOpen
  \bibfield  {author} {\bibinfo {author} {\bibfnamefont {M.}~\bibnamefont
  {Ponmurugan}},\ }\href@noop {} {}\bibinfo {howpublished} {arXiv:1110.5153v2
  [cond-mat.stat-mech]}\BibitemShut {NoStop}%
\bibitem [{\citenamefont {Kurchan}(1998)}]{kurchan}%
  \BibitemOpen
  \bibfield  {author} {\bibinfo {author} {\bibfnamefont {J.}~\bibnamefont
  {Kurchan}},\ }\href@noop {} {\bibfield  {journal} {\bibinfo  {journal} {J.
  Phys. A: Math. Gen.}\ }\textbf {\bibinfo {volume} {31}},\ \bibinfo {pages}
  {3719} (\bibinfo {year} {1998})}\BibitemShut {NoStop}%
\bibitem [{\citenamefont {Sekimoto}(1998)}]{sekimoto1}%
  \BibitemOpen
  \bibfield  {author} {\bibinfo {author} {\bibfnamefont {K.}~\bibnamefont
  {Sekimoto}},\ }\href@noop {} {\bibfield  {journal} {\bibinfo  {journal}
  {Prog. Theor. Phys. Suppl.}\ }\textbf {\bibinfo {volume} {130}},\ \bibinfo
  {pages} {17} (\bibinfo {year} {1998})}\BibitemShut {NoStop}%
\bibitem [{\citenamefont {Sekimoto}(2010)}]{sekimoto2}%
  \BibitemOpen
  \bibfield  {author} {\bibinfo {author} {\bibfnamefont {K.}~\bibnamefont
  {Sekimoto}},\ }\href@noop {} {\emph {\bibinfo {title} {Stochastic
  Energetics}}},\ \bibinfo {series} {Lecture Notes in Physics}, Vol.\ \bibinfo
  {volume} {799}\ (\bibinfo  {publisher} {Springer},\ \bibinfo {year}
  {2010})\BibitemShut {NoStop}%
\bibitem [{\citenamefont {Celani}\ \emph {et~al.}(2012)\citenamefont {Celani},
  \citenamefont {Bo}, \citenamefont {Eichhorn},\ and\ \citenamefont
  {Aurell}}]{Celani12}%
  \BibitemOpen
  \bibfield  {author} {\bibinfo {author} {\bibfnamefont {A.}~\bibnamefont
  {Celani}}, \bibinfo {author} {\bibfnamefont {S.}~\bibnamefont {Bo}}, \bibinfo
  {author} {\bibfnamefont {R.}~\bibnamefont {Eichhorn}}, \ and\ \bibinfo
  {author} {\bibfnamefont {E.}~\bibnamefont {Aurell}},\ }\href@noop {}
  {\bibfield  {journal} {\bibinfo  {journal} {Phys. Rev. Lett.}\ }\textbf
  {\bibinfo {volume} {109}},\ \bibinfo {pages} {260603} (\bibinfo {year}
  {2012})}\BibitemShut {NoStop}%
\bibitem [{\citenamefont {Ge}(2014)}]{Ge14}%
  \BibitemOpen
  \bibfield  {author} {\bibinfo {author} {\bibfnamefont {H.}~\bibnamefont
  {Ge}},\ }\href@noop {} {\bibfield  {journal} {\bibinfo  {journal} {Phys. Rev.
  E}\ }\textbf {\bibinfo {volume} {89}},\ \bibinfo {pages} {022127} (\bibinfo
  {year} {2014})}\BibitemShut {NoStop}%
\bibitem [{\citenamefont {Ford}()}]{Ford15b}%
  \BibitemOpen
  \bibfield  {author} {\bibinfo {author} {\bibfnamefont {I.~J.}\ \bibnamefont
  {Ford}},\ }\href@noop {} {}\bibinfo {howpublished} {arXiv:1508.03488
  [cond-mat.stat-mech]}\BibitemShut {NoStop}%
\bibitem [{\citenamefont {Rohrbach}(2005)}]{Rohrbach05}%
  \BibitemOpen
  \bibfield  {author} {\bibinfo {author} {\bibfnamefont {A.}~\bibnamefont
  {Rohrbach}},\ }\href@noop {} {\bibfield  {journal} {\bibinfo  {journal}
  {Phys. Rev. Lett.}\ }\textbf {\bibinfo {volume} {95}},\ \bibinfo {pages}
  {168102} (\bibinfo {year} {2005})}\BibitemShut {NoStop}%
\bibitem [{\citenamefont {Piazza}\ and\ \citenamefont
  {Parola}(2008)}]{Piazza08}%
  \BibitemOpen
  \bibfield  {author} {\bibinfo {author} {\bibfnamefont {R.}~\bibnamefont
  {Piazza}}\ and\ \bibinfo {author} {\bibfnamefont {A.}~\bibnamefont
  {Parola}},\ }\href@noop {} {\bibfield  {journal} {\bibinfo  {journal} {J.
  Phys.: Condens. Matter}\ }\textbf {\bibinfo {volume} {20}},\ \bibinfo {pages}
  {153102} (\bibinfo {year} {2008})}\BibitemShut {NoStop}%
\bibitem [{\citenamefont {Hatano}\ and\ \citenamefont
  {Sasa}(2001)}]{hatanosasa}%
  \BibitemOpen
  \bibfield  {author} {\bibinfo {author} {\bibfnamefont {T.}~\bibnamefont
  {Hatano}}\ and\ \bibinfo {author} {\bibfnamefont {S.~I.}\ \bibnamefont
  {Sasa}},\ }\href@noop {} {\bibfield  {journal} {\bibinfo  {journal} {Phys.
  Rev. Lett.}\ }\textbf {\bibinfo {volume} {86}},\ \bibinfo {pages} {3463}
  (\bibinfo {year} {2001})}\BibitemShut {NoStop}%
\bibitem [{\citenamefont {Speck}\ and\ \citenamefont
  {Seifert}(2005)}]{IFThousekeeping}%
  \BibitemOpen
  \bibfield  {author} {\bibinfo {author} {\bibfnamefont {T.}~\bibnamefont
  {Speck}}\ and\ \bibinfo {author} {\bibfnamefont {U.}~\bibnamefont
  {Seifert}},\ }\href@noop {} {\bibfield  {journal} {\bibinfo  {journal} {J.
  Phys. A: Math. Gen.}\ }\textbf {\bibinfo {volume} {38}},\ \bibinfo {pages}
  {L581} (\bibinfo {year} {2005})}\BibitemShut {NoStop}%
\bibitem [{\citenamefont {{R.E. Spinney and I.J.
  Ford}}(2012)}]{SpinneyFord12b}%
  \BibitemOpen
  \bibfield  {author} {\bibinfo {author} {\bibnamefont {{R.E. Spinney and I.J.
  Ford}}},\ }\href@noop {} {\bibfield  {journal} {\bibinfo  {journal} {Phys.
  Rev. E}\ }\textbf {\bibinfo {volume} {85}},\ \bibinfo {pages} {051113}
  (\bibinfo {year} {2012})}\BibitemShut {NoStop}%
\end{thebibliography}

%

\end{document}